# Optical properties of graphene antidot lattices

Thomas G. Pedersen[1], Christian Flindt[2], Jesper Pedersen[2], Antti-Pekka Jauho[2,3], Niels Asger Mortensen[2] and Kjeld Pedersen[1]

[1]*Department of Physics and Nanotechnology, Aalborg University, DK-9220 Aalborg Ø, Denmark*
[2]*Department of Micro and Nanotechnology, NanoDTU, Technical University of Denmark, DK-2800 Kongens Lyngby, Denmark*
[3]*Laboratory of Physics, Helsinki University of Technology, P. O. Box 1100, 02015 HUT, Finland*

Undoped graphene is semi-metallic and thus not suitable for many electronic and optoelectronic applications requiring gapped semiconductor materials. However, a periodic array of holes (antidot lattice) renders graphene semiconducting with a controllable band gap. Using atomistic modelling, we demonstrate that this artificial nanomaterial is a dipole-allowed direct gap semiconductor with a very pronounced optical absorption edge. Hence, optical infrared spectroscopy should be an ideal probe of the electronic structure. To address realistic experimental situations, we include effects due to disorder and the presence of a substrate in the analysis.

## 1. Introduction

Graphene has emerged as a promising material for nanoscale electronic devices. Most importantly, graphene combines a high mobility (~15000 cm$^2$/Vs [1, 2]) with the possibility of patterning using e-beam lithography [2-4]. In addition, the very long spin-coherence time is important for potential spintronics applications [5, 6]. Patterning of graphene into Hall bars [1, 7], quantum dots [2,3], nanoribbons [4] and circular Aharonov-Bohm interferometers [8] has been demonstrated. Recently, we have proposed adding graphene antidot lattices [9] to this list. Our proposed antidot structure consists of a hexagonal array of circular holes perforating the graphene sheet. Such a periodic perturbation turns the semi-metallic sheet into a semiconductor with a controllable band gap. Furthermore, "defects" in the lattice formed by leaving one or several unit cells intact support localized states that could lead to realization of a graphene spin qubit architecture [9]. However, the fully periodic antidot lattice is highly interesting in itself. For instance, transport under magnetic fields could lead to Hofstadter butterfly features [10]. Also, the tuneable band gap could be used to design quantum wells and channels for electronic devices. It is even conceivable that tuneable absorption and emission of light could lead to novel graphene optoelectronic devices.

In this work, we present a theoretical study of the optical properties of graphene antidot lattices. Expanding on our previous work [9], we demonstrate how optical spectroscopy will be useful in characterizing the electronic structure of antidot lattices. In particular, we predict a highly visible absorption edge corresponding to the band gap. Hence, optical (infrared) absorption spectroscopy is a promising candidate for characterization. We compute the optical properties using a tight-binding formalism [11]. To accelerate convergence with respect to *k*-point sampling, an improved triangle integration method including *k*-dependent matrix elements has been developed. We present a systematic study of the absorption signature in two different families of lattices as the size of the perforation increases. In practise, variation in hole position and size/shape will lead to inhomogeneously broadened spectra. We study the influence of broadening on the optical spectra to gauge the effect on the measurable response. Also, samples placed on substrates are considered. We find that even in the presence of broadening, both absorption and reflection contrast spectra display clearly detectable band gap features. Finally, the dependence of the low-frequency refractive index on energy gap is analyzed for a large compilation of antidot structures.

## 2. Theory and methods

The optical properties of an extremely thin layer, such as monolayer graphene, can be characterized in two distinct ways. Physically, it is appropriate to view the layer as a charge sheet with complex sheet conductivity $\tilde{\sigma}(\omega)$. Alternatively, the sheet may be viewed as a homogeneous layer with a small but finite thickness $d_g$, taken as the graphite inter-layer lattice constant $\sim 3.35$ Å, and characterized by a dielectric constant $\varepsilon(\omega)$. As long as the layer thickness is much less than the wavelength, the two approaches lead to virtually identical results provided the response functions are related via $\tilde{\sigma}(\omega) = -i d_g \varepsilon_0 \omega [\varepsilon(\omega) - 1]$. The antidot lattice is a periodic structure and as such all properties are calculated as appropriate integrals over a 2-dimensional Brillouin zone. We apply the following approach in all computations: First, the limit of vanishing broadening is considered. This allows us to calculate the real part of the conductivity using a highly accurate triangle integration method including *k*-dependent matrix elements. The details of this method are given in the appendix. Second, the imaginary part of the conductivity is obtained via a Kramers-Kronig transform. Finally, broadening is reintroduced by convoluting with a Gaussian line broadening function. We consider only fully periodic structures and ignore exciton effects in the present work. Localized excitons produce additional discrete absorption resonances below the band gap and the continuous spectrum above the gap is modified by continuum excitons. By ignoring electron-hole interaction, we disregard discrete resonances and approximate the continuous spectrum by the single-electron response. The single-electron response is sufficiently complex and computationally demanding that we choose to postpone exciton effects to future work, however. Following Ref. [12], the real part of the conductivity $\sigma(\omega) = \text{Re}\,\tilde{\sigma}(\omega)$ at low temperature is

$$\sigma(\omega) = \frac{e^2}{2\pi m^2 \omega} \sum_{v,c} \int |P_{vc}|^2 \delta\left(E_{cv}(\vec{k}) - \hbar\omega\right) d^2k, \tag{1}$$

where $P_{vc}$ is the in-plane momentum matrix element and $E_{cv} \equiv E_c - E_v$ is the transition energy between valence band *v* and conduction band *c*. This expression applies to regular graphene as well as graphene antidot lattices provided the integral is taken over appropriate Brillouin zones. Energies and matrix elements are computed from tight-binding eigenstates. We use a simple orthogonal $\pi$-electron model with a nearest-neighbour transfer integral of $\gamma = 3.033$ eV [13]. This model is known to agree with the first-principles band structure in the low-energy range. Corrections for edge effects can be incorporation into the transfer integral but leads only to a slight opening of the antidot band gap [9]. The momentum operator is given solely by the $k$-space gradient of the tight-binding Hamiltonian $\vec{P} = (m/\hbar)\nabla_k H$ since intra-atomic terms are absent in $\pi$-electron models [11].

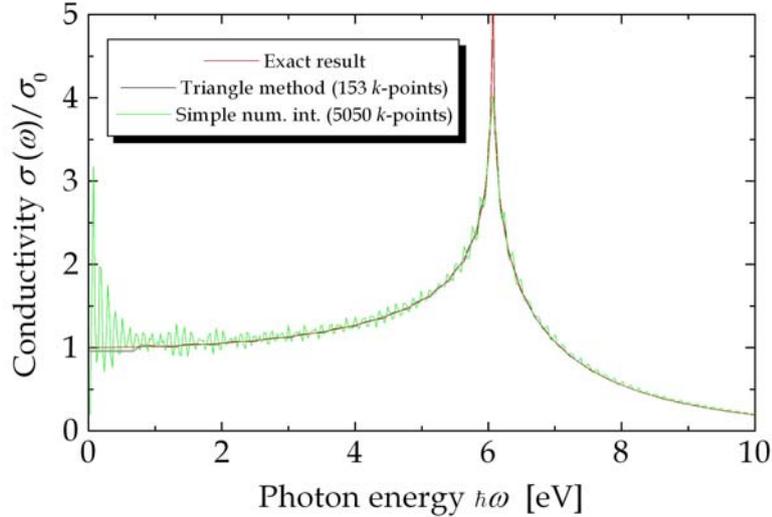

Figure 1. Comparison of numerical triangle integration with the exact conductivity and simple discretization. Note the different $k$-point sampling for the two numerical schemes.

As a reference, we first consider a regular graphene sheet without an antidot lattice. In this case, the analytic results of Ref. [12] (correcting typographical errors) yield a conductivity

$$\sigma(\omega) = \frac{e^2}{\sqrt{24}\pi\hbar\Omega^{3/2}} \mathrm{Re}\left[\frac{144 - 12\Omega + \Omega^3}{24} K\left(\frac{(6-\Omega)(2+\Omega)^3}{128\Omega}\right) - 12 E\left(\frac{(6-\Omega)(2+\Omega)^3}{128\Omega}\right)\right], \quad (2)$$

where $\Omega = \hbar\omega/\gamma$ and $K$ and $E$ are elliptic integrals. It can be shown that taking the zero frequency limit of the above expression leads to a minimum graphene conductivity of $\sigma = \sigma_0 \equiv e^2/4\hbar$ in agreement with several other calculations [14, 15]. Retaining the first non-vanishing correction one finds $\sigma \approx \sigma_0(1 + \Omega^2/9)$ at low frequencies. The low frequency response is modified if the chemical potential is shifted away from the Dirac point via doping [16] but in the present work only intrinsic graphene is considered. To illustrate the accuracy of the improved triangle method we compare in Fig. 1 the exact result given by Eq.(2) to numerical integration based on (a) the triangle method with 153 $k$-points and (b) simple rectangular discretization of Eq.(1) using 5050 $k$-points and a broadening of 20 meV. Numerical integration is taken over the irreducible Brillouin zone using a symmetrised matrix element $|P_{vc}|^2 = (|P_{vc}^x|^2 + |P_{vc}^y|^2)/2$. Even with only a fraction of the *k*-points, the triangle integration is clearly superior to simple discretization. Moreover, the agreement with the exact curve is excellent.

## 3. Results

We now turn to antidot lattices in which an energy gap opens around the Fermi level. As demonstrated in Ref. [9], perforation of a graphene sheet by a regular hexagonal array of circular holes yields a gapped band structure that can be controlled to a large extent by varying the radius and distance between holes. In addition, hole shape may play an important role in determining the properties. For instance, replacing the circular perforation with a triangular one having zig-zag edges produces a dispersionless "metallic" band at the Fermi level. The different band structures are illustrated in Fig. 2. Here, the circular structure is a {10,3} antidot lattice in the {*L*,*R*} notation suggested in Ref. [9]: *L* is the side length and *R* the radius of the perforation, both in units of the graphene lattice constant. The unit cell of the triangular structure is similar in size to the circular case and the area of the triangular perforation is roughly equal to that of the circular hole. The band width of the dispersionless band is identically zero because the simple nearest-neighbour model allows for eigenstates in which the node-structure completely decouples all occupied $\pi$ - orbitals in the zig-zag case. If interactions beyond nearest neighbours are included, a small but finite band width is observed. Antidot lattices with such triangular perforations would lead to additional interesting features in the optical response such as controllable transparency windows. Presumably, their fabrication using e.g. e-beam lithography will be rather demanding, however, and in the remaining part of the paper we focus on circular perforations.

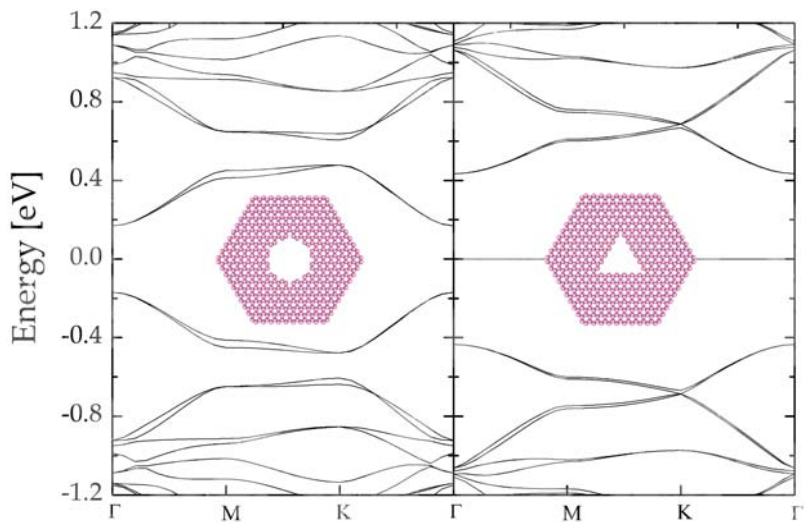

Figure 2. Band structures of a {10,3} antidot lattice and similar structure having a triangular hole with zig-zag edges. Notice the dispersionless band at 0 eV in the triangular case.

For numerical integration, the irreducible Brillouin zone is partitioned into 4098 triangles, which is equivalent to 2145 unique *k*-points. In usual two-dimensional direct band gap semiconductors with parabolic energy dispersion, the absorption edge is a clearly discernable step profile [17]. Our numerical results show that a similar behaviour is found in graphene antidot lattices. As an illustration, in Figs. 3 and 4 the conductivity spectra are shown for $\{10,R\}$ and $\{12,R\}$ lattices The step-like absorption edge coincides with the band gap and demonstrates that antidot lattices are two-dimensional dipole-allowed direct gap semiconductors. This will be important for possible optoelectronic applications including light emission and detection. Also, experimental detection of band gaps using infrared spectroscopy should be feasible with such a clear signature.

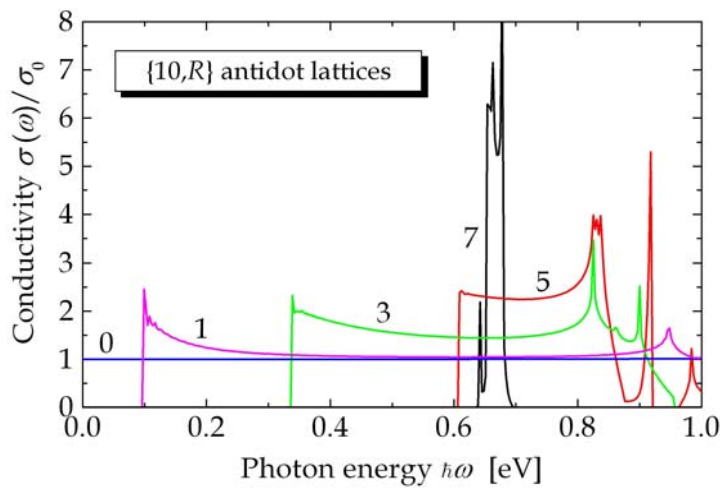

Figure 3. Conductivity spectra for several $\{10,R\}$ antidot lattices with *R* indicated next to each spectrum. The conductivity is normalized to the DC value $\sigma_0 = e^2/4\hbar$.

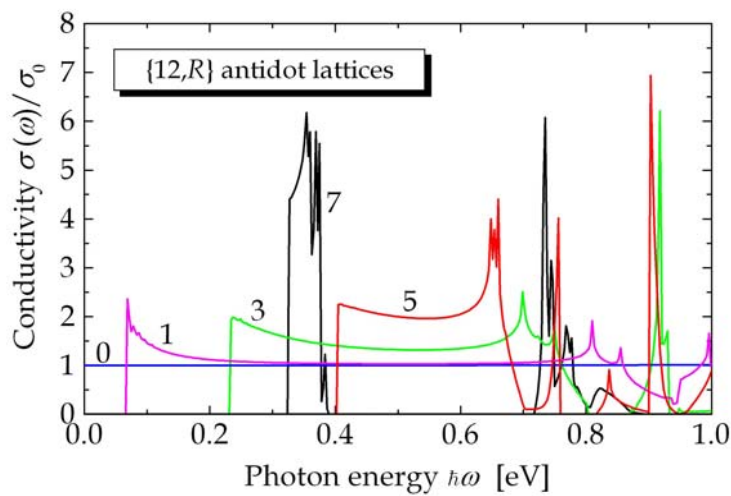

Figure 4. Same as Fig. 3 but for the $\{12,R\}$ family of lattices.

To fully characterize the optical properties we need to determine both real and imaginary parts of the frequency-dependent conductivity. Also, inhomogeneous broadening must be considered as practical e-beam patterning will lead to variations in hole size, shape and position. The imaginary part of $\tilde{\sigma}(\omega)$ is readily obtained from a Kramers-Kronig transform of the real part. Care should be taken, however, that the real part is calculated up to sufficiently large frequencies. Subsequently, broadening can be included by convoluting with a Gaussian line shape function $\exp[-(\omega-\omega')^2/\Gamma^2]/(\Gamma\sqrt{\pi})$. The broadening $\Gamma$ reflects the degree of disorder and we estimate that high quality samples should have $\hbar\Gamma <$ 100 meV. In Fig. 5, we show the effect on the complex conductivity of broadening by $\hbar\Gamma =$ 20 meV and 50 meV. Increased broadening tends to blur finer features in the spectra but at this level of disorder the absorption edge is still clearly visible. We emphasize that it is the step-like absorption edge of the two-dimensional semiconductor that makes band edge detection feasible for samples with relatively low disorder.

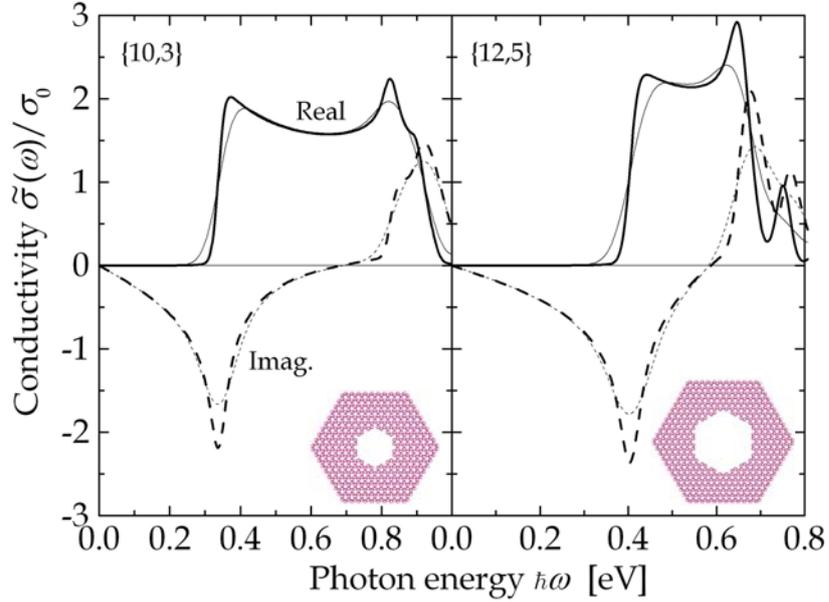

Figure 5. Complex conductivity spectra including broadening. Curves are real parts (solid lines) and imaginary parts (dashed lines) for $\hbar\Gamma =$ 20 meV (thick lines) and $\hbar\Gamma =$ 50 meV (thin lines).

In practical experiments, graphene samples are usually positioned on a suitable substrate for investigations. Hence, it is of importance to discuss the role of substrates on the optical signatures of antidot lattices. For transmission measurements any transparent substrate can be used and the recorded spectrum will essentially provide the real part of the conductivity directly. Alternatively, a reflection geometry can be used. Usually, an oxidized silicon wafer is applied for this purpose. In fact, monolayer graphene is usually identified in mechanically peeled graphite flakes by observing flakes of varying thickness on oxidized Si wafers in an optical microscope [1, 2]. Using white light illumination and an oxide thickness of 300 nm it turns out that even monolayer graphene is clearly visible in the microscope. The contrast, which is around 15%, is a result of a fortuitous choice of oxide thickness and the large conductivity of graphene [18-20]. The geometry of such a sample is illustrated in Fig. 6. We denote the frequency dependent refractive indices of SiO$_2$ and Si by $n_1$ and $n_2$, respectively, and the oxide thickness by $d$. Introducing a dimensionless graphene conductivity $\bar{\sigma} \equiv \tilde{\sigma}/\varepsilon_0 c$ the reflectance at normal incidence is given by

$$R = \left| \frac{e^{2idn_1\omega/c}(1+n_1-\bar{\sigma})(n_1-n_2)+(1-n_1-\bar{\sigma})(n_1+n_2)}{e^{2idn_1\omega/c}(1-n_1+\bar{\sigma})(n_1-n_2)+(1+n_1+\bar{\sigma})(n_1+n_2)} \right|^2 . \quad (3)$$

A general expression valid at arbitrary angle of incidence is given in Ref. [20]. The reflectance contrast is defined as $(R_0 - R)/R_0$, where $R_0$ is calculated as above but taking $\bar{\sigma} = 0$. We take experimental refractive indices of SiO$_2$ and Si from Refs. [21] and [22], respectively.

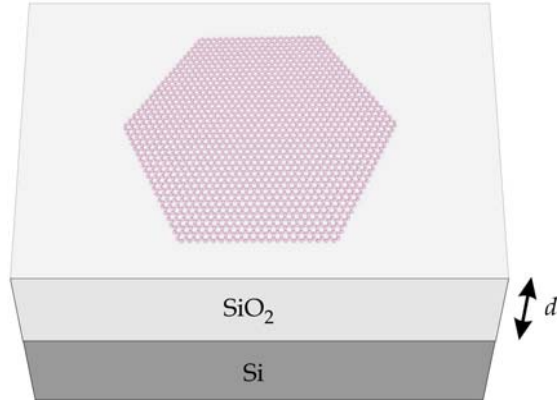

Figure 6. Graphene sample positioned on an oxidized Si wafer.

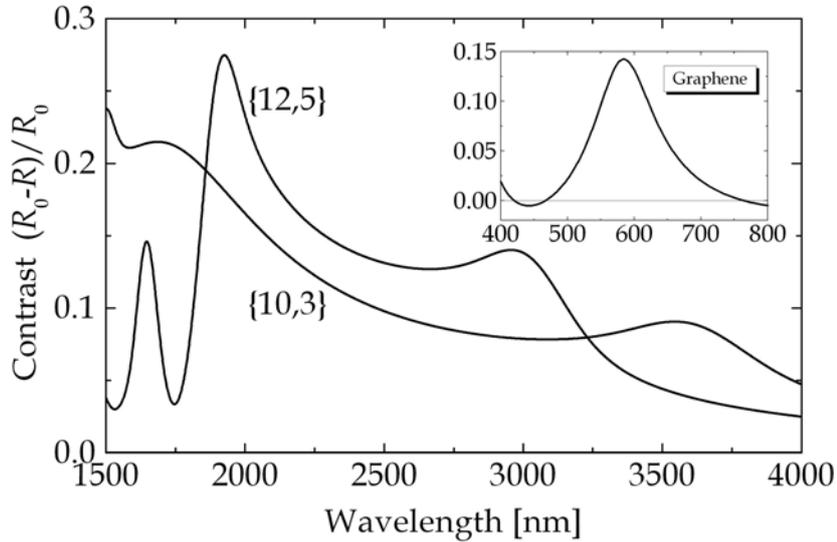

Figure 7. Infrared reflectance contrast for {10,3} and {12,5} lattices. Inset: contrast for regular graphene in the visible.

In Fig. 7, we have displayed the reflectance contrast of {10,3} and {12,5} antidot lattices positioned on 300 nm oxide Si wafers. In the computation, the complex conductivity spectra shown in Fig. 4 for the case $\hbar\Gamma$ = 20 meV have been applied. It is apparent that a large contrast exceeding 20% is predicted for this situation. In the inset, the contrast for regular graphene in the visible is illustrated for comparison. The magnitude and location of the resonance just below 600 nm is in good agreement with experiments and other calculations [18]. For the {10,3} and {12,5} structures the band gaps are around 0.34 eV and 0.41 eV, respectively, c.f. Figs. 3 and 4. This corresponds to resonance wavelengths of 3650 nm and 3025 nm. These resonances are clearly observed in the contrast plots in Fig. 7. Hence, reflectance contrast measurements could be a viable method of determining band edges. Recently, optical spectroscopy on gated graphene in precisely this wavelength range has been reported [23], which further supports the feasibility of our proposal.

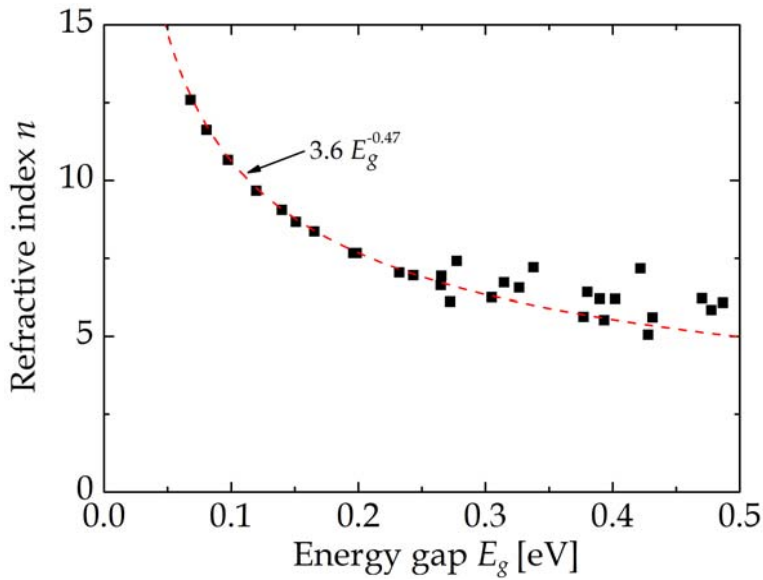

Figure 8. Refractive index at low frequencies vs. energy gap for different antidot lattices. The dashed line is a power law fit to the data.

Several optical and electro-optic applications of graphene antidot lattices can be envisioned. For instance, light emitting devices tailored to specific wavelengths could be fabricated. It might also be possible to incorporate antidot lattices into wave guiding structures fabricated on e.g. the $SiO_2$ substrate. For all optical and electro-optic applications, the refractive index $n$ is of importance and we wish to study the effect of antidot geometry on $n$. To this end, we compute the complex refractive index via the relation $\tilde{n}(\omega) = \sqrt{1 + i\tilde{\sigma}(\omega)/(d_g \varepsilon_0 \omega)}$ for a compilation of different antidot structures with $L$ in the range from 4 to 12 including both large and small energy gap structures. Below the gap $E_g$, the real part of the refractive index dominates and we focus on the real-valued low-frequency $n = \tilde{n}(0)$ limit. In semiconductors, the refractive index generally decreases with increasing energy gap because more remote electronic transitions make little contribution at low frequencies. A similar tendency is observed in graphene antidot samples, as illustrated in Fig. 8. In the range of small energy gaps, $n$ scales approximately as a power law $E_g^{-0.47}$. Hence, tuneability of the optical properties also includes the refractive index. The attainable values become very large for low energy gap structures approaching the behaviour of unmodified graphene.

## 4. Summary

In summary, the optical response of graphene antidot lattices has been analyzed with a $\pi$-electron tight-binding model. We find that these structures behave as dipole-allowed direct gap two-dimensional semiconductors. The optical properties have been computed using an improved triangle method capable of handling large structures with great accuracy. In addition, inhomogeneous broadening caused by disorder is taken into account. We find that optical infrared spectroscopy is ideally suited for probing the electronic structure. Placing the antidot sample on a dielectric substrate, the response can be probed in both reflection and transmission geometries. We predict clearly visible band gap features in both modes. Finally, the low-frequency refractive index has been studied for a range of different antidot geometries and we find that the refractive index follows a decreasing power-law behaviour with increasing energy gap.

## Appendix: Improved triangle method

The triangle method [24] of approximating two-dimensional integrals of resonant functions is similar to the well-known 3D tetrahedron method. Key to the method is a linearization of *k*-dependent energies inside small triangular sections of the Brillouin zone. We demonstrate in this appendix that it is possible to include *k*-dependent matrix elements and, thereby, increase the accuracy of response function calculations. We consider an integral of the form

$$S(\omega) = \int F(\vec{k})\delta(E_{cv}(\vec{k}) - \hbar\omega)d^2k \\ = \sum_{\triangle} \int_{\triangle} F(\vec{k})\delta(E_{cv}(\vec{k}) - \hbar\omega)d^2k. \quad (A.1)$$

Here, "$\triangle$" denotes a triangle and the sum is over a triangular mesh covering the (irreducible) Brillouin zone. The integral can be reduced to a line integral along the curve $l(\triangle)$ on which $E_{cv}(\vec{k}) = \hbar\omega$ according to

$$S(\omega) = \sum_{\triangle} \int_{l(\triangle)} \frac{F(\vec{k})}{|\nabla_k E_{cv}(\vec{k})|} dl \approx \sum_{\triangle} \frac{1}{|\nabla_k E_{cv}|} \int_{l(\triangle)} F(\vec{k}) dl, \quad (A.2)$$

where the linear approximation for $E_{cv}(\vec{k})$ has been assumed. This approximation means further that $l(\triangle)$ becomes a straight line as illustrated in Fig. 9.

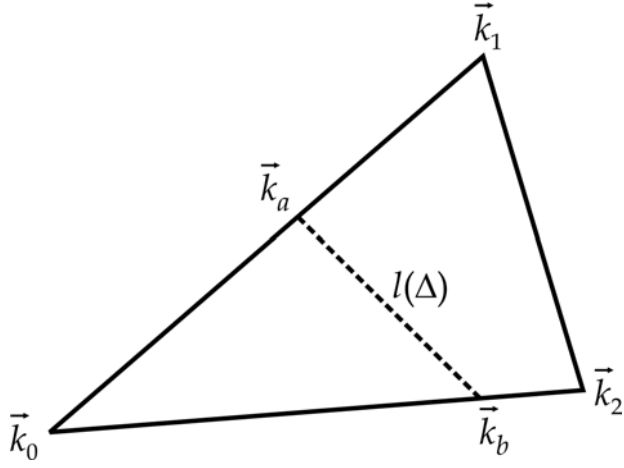

Figure 9. Important *k*-points used in the improved triangle method.

We now invoke the linear approximation for $F(\vec{k})$ as well. Hence, the remaining task is reduced to integrating a linearly varying function along a straight line. We introduce the compact notation $E_i \equiv E_{cv}(\vec{k}_i)$ and take the transition energies in the three corners to be ordered according to $E_0 \leq E_1 \leq E_2$. Provided $E_0 \leq \hbar\omega < E_1$, start ($\vec{k}_a$) and end ($\vec{k}_b$) points of the line $l(\Delta)$ are located at (c.f. Fig. 9)

$$\vec{k}_a = \vec{k}_0 + (\vec{k}_1 - \vec{k}_0)\frac{\hbar\omega - E_0}{E_{10}}, \quad \vec{k}_b = \vec{k}_0 + (\vec{k}_2 - \vec{k}_0)\frac{\hbar\omega - E_0}{E_{20}}, \quad (A.3)$$

with $E_{ij} \equiv E_i - E_j$. A slightly different expression for $\vec{k}_a$ is found for the case $E_1 \leq \hbar\omega < E_2$. Hence, the integral in Eq.(A.2) becomes

$$S(\omega) \approx \sum_\Delta \frac{l(\Delta)}{2|\nabla_k E_{cv}|}\left[F(\vec{k}_a) + F(\vec{k}_b)\right]. \quad (A.4)$$

Using simple algebra, all quantities may be expressed in terms of values in the three corners and the triangle area $A_\Delta$ and we can finally write $S(\omega) \approx \sum_\Delta S_\Delta(\omega)$ with

$$S_\Delta(\omega) = 2A_\Delta \begin{cases} \dfrac{\hbar\omega - E_0}{E_{10}E_{20}}\left[F_0 + \dfrac{\hbar\omega - E_0}{2}\left(\dfrac{F_{10}}{E_{10}} + \dfrac{F_{20}}{E_{20}}\right)\right], & E_0 \leq \hbar\omega < E_1 \\ \dfrac{E_2 - \hbar\omega}{E_{21}E_{20}}\left[F_2 + \dfrac{\hbar\omega - E_2}{2}\left(\dfrac{F_{21}}{E_{21}} + \dfrac{F_{20}}{E_{20}}\right)\right], & E_1 \leq \hbar\omega < E_2. \end{cases} \quad (A5)$$

This expression allows us to evaluate resonant integrals using relatively few *k*-points and retaining great accuracy.